\documentclass[10pt,a4paper]{article}
\usepackage[T1]{fontenc}
\usepackage[utf8]{inputenc}
\usepackage[margin=0.72in]{geometry}
\usepackage{amsmath,amssymb,bm}
\usepackage{mathptmx}
\usepackage{graphicx}
\usepackage{dblfloatfix}

\usepackage{booktabs,multirow,array,xcolor}
\usepackage{caption}
\usepackage{float}
\usepackage[hidelinks]{hyperref}
\usepackage{microtype}
\DeclareCaptionLabelSeparator{bar}{\space|\space}
\newcommand{\panelletter}[1]{{\sffamily\bfseries #1}}
\begin{document}

\twocolumn[{
\begin{center}
{\Large\bfseries Accurately Generalizing Deep-Learning Hamiltonian Methods to Plane-Wave Density Functional Theory\par}
\medskip

Chuntian Wu\textsuperscript{1}, Jiaxiang Li\textsuperscript{1}, Qiang Xu\textsuperscript{1,\ensuremath{\dagger}}, Xuecheng Shao\textsuperscript{1,2,3,\ensuremath{\dagger}}, Congwei Xie\textsuperscript{4,\ensuremath{\dagger}}, Yu Xie\textsuperscript{1,6,\ensuremath{\dagger}}, and Yanming Ma\textsuperscript{1,2,3,6,7,\ensuremath{\dagger}}\par
\emph{\textsuperscript{1} Key Laboratory of Material Simulation Methods \& Software of Ministry of Education, College of Physics, Jilin University, Changchun 130012, China\\
\textsuperscript{2} State Key Laboratory of High Pressure and Superhard Materials, College of Physics, Jilin University, Changchun 130012, China\\
\textsuperscript{3} International Center of Future Science, Jilin University, Changchun 130012, China\\
\textsuperscript{4} Research Center for Crystal Materials, State Key Laboratory of Functional Materials and Devices for Special Environmental Conditions, Xinjiang Key Laboratory of Functional Crystal Materials, Xinjiang Technical Institute of Physics and Chemistry, Chinese Academy of Sciences, Urumqi, China\\
\textsuperscript{5} Key Laboratory of Physics and Technology for Advanced Batteries (Ministry of Education), College of Physics, Jilin University, Changchun 130012, China\\
\textsuperscript{6} School of Physics and Institute of Fundamental and Transdisciplinary Research, Zhejiang University, Hangzhou 310027, China\\
\textsuperscript{7} Center for High-Pressure Science and Technology, Zhejiang University, Hangzhou 310027, China\\
\textsuperscript{\ensuremath{\dagger}} {\small Corresponding authors: \texttt{xuq@jlu.edu.cn}; \texttt{cwxie@ms.xjb.ac.cn}; \texttt{xieyu@jlu.edu.cn}; \texttt{shaoxc@jlu.edu.cn}; \texttt{mym@jlu.edu.cn}}}\par
(Dated: October 13, 2026)

\end{center}
\begin{center}
\begin{minipage}{0.85\textwidth}
\small
\noindent Deep neural networks that predict density functional theory (DFT) Hamiltonians from material structures offer a route to efficient electronic-structure calculations. However, extending localized atomic-orbital (AO) Hamiltonian learning to plane-wave (PW) calculations with hybrid functionals requires a compact AO representation that preserves the effects of nonlocal operators. Here we propose a projectability-guided construction of nonorthogonal AO Hamiltonians from a small, selected set of PW eigenstates and eigenvalues. High-projectability states are retained, and selected higher-energy states complete the AO subspace, avoiding an empirical shift and its arbitrary contribution to the learning targets. Because the construction uses PW eigenstates and eigenvalues, it incorporates nonlocal contributions present in the reference calculation without explicitly reconstructing their AO matrix elements. Benchmarks for MoS\textsubscript{2} and sodium phases show accurate reproduction of PW bands with smaller AO bases than real-space reconstruction. The resulting PBE and HSE Hamiltonians can be learned by equivariant neural networks, connecting PW reference calculations with electronic-structure prediction. This approach broadens the range of PW DFT calculations available for AO Hamiltonian learning and enables compact, accurate training targets for developing more widely applicable electronic-structure models.
\end{minipage}
\end{center}
\vspace{0.5em}
}]

Density functional theory (DFT) is widely used to compute the electronic structure of crystalline materials \cite{ref1,ref2}. Repeated self-consistent calculations, however, become costly for large structures and extensive structural datasets \cite{ref3}. Deep-learning electronic Hamiltonian method can avoid it while yielding band structures and wave functions through diagonalization, from which a range of electronic properties can be calculated \cite{ref4,ref5}. Frameworks such as DeepH-E3 \cite{ref6}, Uni-HamGNN \cite{ref7} and MACE-H \cite{ref8} have achieved accurate predictions of DFT Hamiltonians expressed in non-orthogonal Linear Combination of Atomic Orbitals (LCAO) bases using E(3)-equivariant graph neural networks. Building on DeepH-E3, DeepH-hybrid has extended this approach to hybrid-functional Hamiltonians \cite{ref9}. MACE-H builds on higher-order equivariant message passing \cite{ref10}. Hamiltonian learning has also been extended to magnetic superstructures \cite{ref11} and, using Hamiltonian derivatives, perturbation-response calculations \cite{ref12}. Together, these developments have made accurate, transferable Hamiltonian models an increasingly practical tool for electronic-structure research \cite{ref13,ref14}.

E(3)-equivariant models such as DeepH-E3 \cite{ref6}, Uni-HamGNN \cite{ref7} and MACE-H \cite{ref8} have achieved accurate predictions of DFT Hamiltonians in localized, nonorthogonal linear-combination-of-atomic-orbitals (LCAO) bases. DeepH-hybrid has further extended this approach to hybrid-functional Hamiltonians \cite{ref9}. Localized AO descriptions provide a compact alternative for DFT \cite{ref15,ref16,ref17}. Many reference calculations for periodic solids, however, are performed in the delocalized PW basis \cite{ref18}, whereas these models learn sparse AO matrix blocks \cite{ref19,ref20}. Earlier conversion routes include projection into AO bases \cite{ref21,ref22} and direct evaluation of AO matrix elements \cite{ref23}. Gong et al. \cite{ref24} connected the two representations by reconstructing AO Hamiltonians from PW DFT results in real space and using them to train DeepH-E3.

In that reconstruction, the kinetic-energy and nonlocal pseudopotential terms are evaluated with two-center integrals, and the local effective potential is integrated on a real-space grid \cite{ref24,ref25}. As formulated, it omits additional terms such as exact exchange in screened hybrid functionals \cite{ref26,ref27} and the Hubbard correction in DFT+U \cite{ref28}. Projection from PW eigenstates can reflect the effects of these terms, but including poorly represented states can introduce spurious bands and distort bands of interest \cite{ref29}. Agapito et al. \cite{ref29} addressed this problem by filtering low-projectability states and shifting the complementary AO subspace. The shift changes the real-space matrix elements used as learning targets, and a value chosen for the training structures cannot be guaranteed to keep the artificial bands outside the energy window of interest in an unseen structure.

Here we construct compact AO Hamiltonians from a projectability-guided selection of PW eigenstates. We complete the AO subspace with selected higher-energy PW states and their eigenvalues, avoiding an empirical shift while reflecting the effects of nonlocal contributions present in the PW calculation. In our benchmarks, the resulting Hamiltonians reproduce PW DFT bands more accurately than the methods evaluated here with fewer AO basis functions, reducing the cost of diagonalization. The construction accommodates both semilocal and hybrid-functional calculations and produces real-space matrices suitable for equivariant Hamiltonian learning (Fig. 1).

\begin{figure*}[tbp]
\centering
\includegraphics[width=0.9\linewidth]{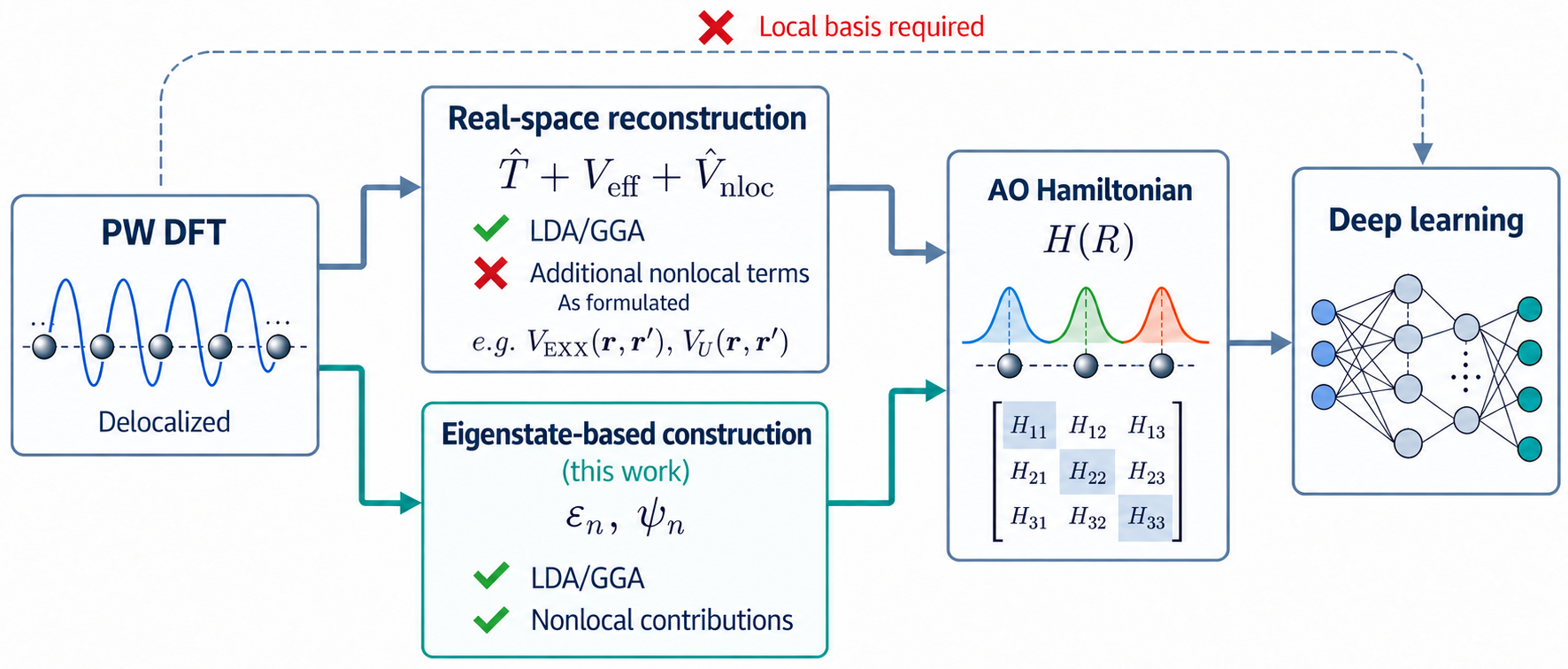}
\caption{\textbf{Routes from PW DFT to localized AO Hamiltonians for machine learning.} Plane waves are delocalized, whereas AO Hamiltonian-learning frameworks use localized matrix blocks. Real-space reconstruction \cite{ref24} includes kinetic energy, the local effective potential and the nonlocal pseudopotential, but its present formulation omits additional nonlocal contributions, such as exact exchange in hybrid functionals and Hubbard corrections in DFT+U. Our construction uses selected PW eigenstates and eigenvalues to reflect the nonlocal contributions included in the reference calculation and produce compact AO Hamiltonians for deep learning.}
\label{fig:1}
\end{figure*}

\section*{Results}

\subsection*{Theory}

The real-space reconstruction reported by Gong \emph{et al.} \cite{ref24} efficiently converts PW DFT results to AO Hamiltonians. Their formulation gives the real-space matrix element as:

\begin{equation}
\begin{aligned}
H_{i\alpha,j\beta} &= \langle\phi_{i\alpha}| - \frac{1}{2}\nabla^{2}|\phi_{j\beta}\rangle \\
& + \int{d^{3}r\phi_{i\alpha}^{\ast}(\mathbf{r})V_{\text{eff}}(\mathbf{r})\phi_{j\beta}(\mathbf{r})} + \langle\phi_{i\alpha}|{\widehat{V}}_{\text{nloc}}|\phi_{j\beta}\rangle.
\end{aligned}
\label{eq:1}
\end{equation}

Here, \(i,j\) label atoms and \(\alpha,\beta\) label their orbitals. The three terms represent the kinetic energy, local effective potential and nonlocal pseudopotential \cite{ref30}, respectively. The first and last are evaluated using two-center integrals \cite{ref15,ref24,ref25}; the local-potential term is integrated on a real-space grid, with \(V_{\text{eff}}\left( \mathbf{r} \right) = V_{\text{Har}}\left( \mathbf{r} \right) + V_{\text{xc}}\left( \mathbf{r} \right) + V_{\text{loc}}\left( \mathbf{r} \right)\) comprising the Hartree, exchange--correlation and local pseudopotential contributions. This formulation omits additional nonlocal terms, including exact exchange \(V_{\text{EXX}}\left( \mathbf{r},\mathbf{r}' \right)\) in hybrid functionals \cite{ref26,ref27} and the Hubbard correction \(V_{\text{U}}\left( \mathbf{r},\mathbf{r}' \right)\) in DFT+U \cite{ref28}, preventing complete reconstruction when these terms are required (Fig. 1).

These additional contributions are already reflected in the PW eigenstates and eigenvalues. Direct projection uses their spectral decomposition to construct an AO Hamiltonian \cite{ref21,ref22,ref29}. Retaining \(N\) eigenstates at a given \(\mathbf{k}\)-point gives:

\begin{equation}
\begin{aligned}
{\overline{H}}_{\alpha\beta}(\mathbf{k}) &= \langle\phi_{\alpha\mathbf{k}}|\widehat{H}|\phi_{\beta\mathbf{k}}\rangle \\
&\approx \sum_{n = 1}^{N}\langle\phi_{\alpha\mathbf{k}}|\psi_{n\mathbf{k}}\rangle E_{n\mathbf{k}}\langle\psi_{n\mathbf{k}}|\phi_{\beta\mathbf{k}}\rangle.
\end{aligned}
\label{eq:2}
\end{equation}

Here, \(E_{nk}\) and \(\psi_{nk}\) are the PW eigenvalue and eigenstate, respectively. The accuracy of the truncated projection depends on the retained states and their representation in the AO basis, measured by the projectability \cite{ref21,ref22}. For a nonorthogonal basis:

\begin{equation}
\begin{aligned}
p_{n\mathbf{k}} &= {\mathbf{b}_{n\mathbf{k}}}^{\dagger}S^{- 1}(\mathbf{k})\mathbf{b}_{n\mathbf{k }}, \\
\left\lbrack \mathbf{b}_{n\mathbf{k}} \right\rbrack_{\alpha} &= \langle\phi_{\alpha\mathbf{k}}|\psi_{n\mathbf{k}}\rangle,
\end{aligned}
\label{eq:3}
\end{equation}

where \(S_{\alpha\beta}\left( \mathbf{k} \right) = \langle\phi_{\alpha\mathbf{k}}|\phi_{\beta\mathbf{k}}\rangle\) is the AO overlap matrix. For normalized PW states, projectability ranges from 0 to 1; values near 1 indicate accurate representation in the AO subspace. Supplementary Fig. 1 illustrates the limited set of well-represented states in monolayer MoS\textsubscript{2}. Agapito \emph{et al.} \cite{ref29} showed that including low-projectability states can cause spurious hybridization and distort high-projectability bands (Fig. xx in SI).

Their construction retains and normalizes high-projectability projections, then shifts the resulting null subspace \cite{ref29}. In the nonorthogonal AO representation, let \(A\left( \mathbf{k} \right)\) contain the normalized overlap vectors \(\frac{\mathbf{b}_{nk}}{\sqrt{p_{nk}}}\). For linearly independent columns, the complementary projector and shifted Hamiltonian are:

\begin{equation}
Q_{N}(\mathbf{k}) = S(\mathbf{k}) - A(\mathbf{k})\left\lbrack A^{\dagger}(\mathbf{k})S^{- 1}(\mathbf{k})A(\mathbf{k}) \right\rbrack^{- 1}A^{\dagger}(\mathbf{k})
\label{eq:4}
\end{equation}

\begin{equation}
{\overline{H}}_{\kappa}(\mathbf{k}) = A(\mathbf{k})E(\mathbf{k})A^{\dagger}(\mathbf{k}) + \kappa Q_{N}(\mathbf{k}).
\label{eq:5}
\end{equation}

Here, \(E\left( \mathbf{k} \right)\) is diagonal in the retained PW eigenvalues, \(Q_{N}\left( \mathbf{k} \right)\) is the AO matrix of the complementary projector, and \(\kappa\) sets the shift energy. For Hamiltonian learning, \(\kappa Q_{N}\) introduces an arbitrary contribution to real-space training targets. A shift chosen for training structures also cannot guarantee that artificial complementary bands remain outside the energy window of interest in unseen structures. Our construction instead completes the AO subspace using selected PW states and their eigenvalues.

Specifically, we retain low-energy PW states with high projectability onto the AO basis. Without orthogonalizing these retained states, we remove their span from candidate higher-energy PW projections and select linearly independent residuals to complete the AO basis. The complementary subspace is assigned the eigenvalues of the selected PW states rather than an arbitrary shift.

Let \(C_{L}\left( \mathbf{k} \right)\) collect the AO expansion coefficients \(\mathbf{c}_{n\mathbf{k}}\) = \(S^{- 1}\left( \mathbf{k} \right)\mathbf{b}_{n\mathbf{k}}\) of the retained states. Suppressing the common \(\mathbf{k}\) argument on the right-hand side, the S-metric projector and the residual of a candidate higher-energy state are

\begin{equation}
\begin{aligned}
P_{L}\left( \mathbf{k} \right) &= C_{L}\left( {C_{L}}^{\dagger}SC_{L} \right)^{- 1}{C_{L}}^{\dagger}S \\
{\widetilde{\mathbf{c}}}_{m\mathbf{k}} &= \left( I - P_{L} \right)\mathbf{c}_{m\mathbf{k}}.
\end{aligned}
\label{eq:6}
\end{equation}

The index m labels a candidate higher-energy PW state. The inverse requires linearly independent retained projections. Each residual is S-orthogonal to their span, while the retained high-projectability projections remain unchanged.

Let \(C_{H}\left( \mathbf{k} \right)\) collect the selected linearly independent residuals, completing \(C_{L}\left( \mathbf{k} \right)\) to the dimension of the AO basis. The diagonal matrices \(E_{L}\left( \mathbf{k} \right)\) and \(E_{H}\left( \mathbf{k} \right)\) contain the PW eigenvalues of the retained low-energy and selected higher-energy states, respectively. With \(C\left( \mathbf{k} \right) = \lbrack C_{L}\left( \mathbf{k} \right),C_{H}\left( \mathbf{k} \right)\rbrack\) and \(E\left( \mathbf{k} \right) = diag\left( E_{L},E_{H} \right)\), the AO Hamiltonian is

\begin{equation}
H\left( \mathbf{k} \right) = S\left( \mathbf{k} \right)C\left( \mathbf{k} \right)E\left( \mathbf{k} \right)C^{\dagger}\left( \mathbf{k} \right)S\left( \mathbf{k} \right).
\label{eq:7}
\end{equation}

On a complete, uniformly weighted mesh \(\mathcal{K}\) of \(N_{k}\) k-points, the real-space Hamiltonian and overlap matrices follow from the inverse Bloch transform \cite{ref18}:

\begin{equation}
H\left( \mathbf{R} \right) = \frac{1}{N_{k}}\sum_{\mathbf{k} \in \mathcal{K}}e^{- 2\pi i\mathbf{k} \cdot \mathbf{R}}H\left( \mathbf{k} \right)
\label{eq:8}
\end{equation}

Here, \(\mathbf{k}\) is in fractional reciprocal coordinates, and \(\mathbf{R}\) is an integer lattice translation. The resulting real-space blocks \(H\left( \mathbf{R} \right)\) provide the AO representation used for subsequent Hamiltonian learning.

In this work, the AO basis functions are required to be separable into radial and angular parts. Their ability to represent PW eigenstates is quantified by the projectability, which can be improved by enlarging the basis or optimizing the radial functions of the existing basis \cite{ref15,ref17,ref22,ref31,ref32}. The AO basis should also be compatible with the pseudopotentials used in the PW calculations \cite{ref17,ref24}.

\subsection*{Hamiltonian construction benchmarks}

We first evaluate the accuracy of the constructed AO Hamiltonians by comparing their band structures, density of states and optical responses with PW DFT references. We also examine the computational costs of Hamiltonian construction and subsequent diagonalization. These benchmarks span monolayer MoS\textsubscript{2}, sodium phases under pressure, an Al surface slab and MgO with an oxygen vacancy.

\subsubsection*{Application to monolayer MoS\textsubscript{2}}

We compare direct eigenstate projection, real-space reconstruction and our method for monolayer MoS\textsubscript{2}. For PBE \cite{ref33} calculations, we use two AO basis sets: Mo:2s1p1d, S:1s1p and Mo:3s2p2d, S:3s2p1d. Supplementary Fig. 1a compares their projectabilities as a function of band index. For the states shown, both bases represent the first 17 PW bands with projectabilities close to unity. The larger basis improves the representation of these states, particularly where the smaller basis shows modest reductions in projectability. For bands 18--20, enlarging the basis substantially increases projectability, but the values remain well below unity. Thus, the near-unity projectability region remains confined to the first 17 bands in this comparison, motivating the selection of well-represented PW states for Hamiltonian construction.

Figure 2b compares the PBE band structures obtained from the two AO Hamiltonian constructions with the PW DFT reference. Real-space reconstruction uses the larger Mo:3s2p2d, S:3s2p1d basis, whereas our method uses the smaller Mo:2s1p1d, S:1s1p basis. Despite its smaller basis, our construction closely follows the PW reference over the energy window shown. Visible deviations remain for real-space reconstruction in some conduction bands, illustrating that a larger AO basis alone does not ensure closer agreement with the PW electronic structure.

For HSE \cite{ref26} calculations, Supplementary Fig. 1b shows the corresponding projectabilities for the same two AO bases. Both bases again provide high projectability for the first 17 bands, whereas the representation of bands 18--20 is substantially poorer, particularly for the smaller basis. Enlarging the basis improves these higher-energy states but does not bring their projectabilities close to unity. Figure 2c compares the HSE band structures using the same basis assignments as in Fig. 2b. Our construction with the smaller basis closely follows the PW DFT reference, whereas real-space reconstruction with the larger basis exhibits substantial deviations, particularly in the conduction bands. This discrepancy arises from the omission of the nonlocal exact-exchange contribution in the reported real-space formulation. The band-energy mean absolute errors (MAEs) over bands 0--17 are summarized in Table 1.

\begin{table}[H]
\refstepcounter{table}
\label{tab:1}
\centering
\includegraphics[width=0.9\linewidth]{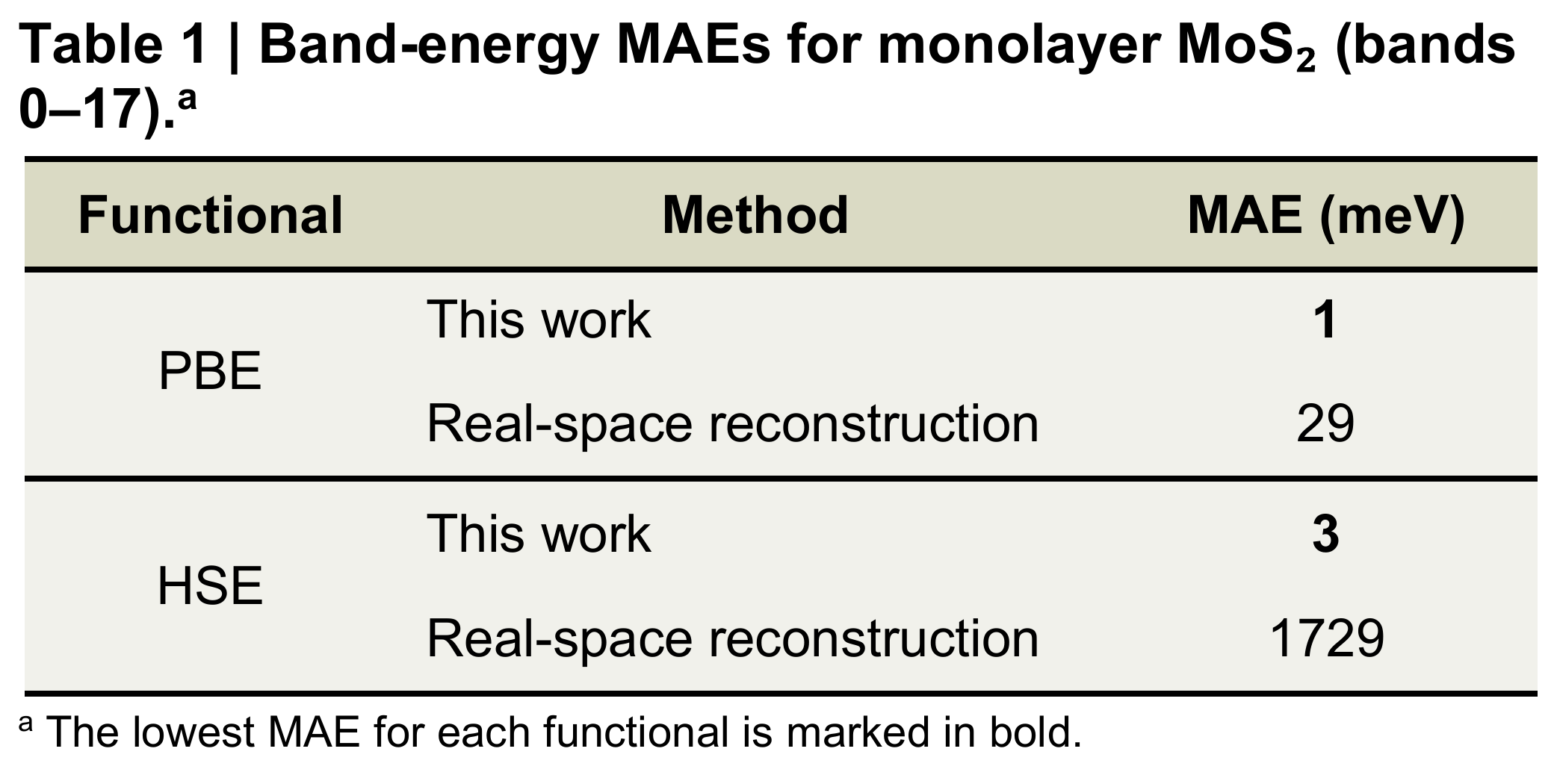}
\end{table}

We also compare the density of states (DOS) of the constructed AO Hamiltonians with the PW DFT reference (Fig. 2d). Our method uses Mo:2s1p1d, S:1s1p, whereas real-space reconstruction uses the larger Mo:3s2p2d, S:3s2p1d basis. With each spectrum referenced to its own midgap, our construction closely reproduces the principal DOS peak positions and amplitudes within the target energy window of \ensuremath{-}7 to 4 eV. The normalized root-mean-square error in this interval is 2.93\% for our method and 14.02\% for real-space reconstruction. Several prominent valence peaks from real-space reconstruction shift towards lower energies, producing larger differences from the PW reference. This comparison is restricted to the target window represented by high-projectability states; higher-energy states with insufficient projectability are excluded from the accuracy assessment.

To assess the reconstructed AO Hamiltonians beyond band dispersions, we calculate the in-plane complex electric susceptibility of monolayer MoS\textsubscript{2} at the independent-particle PBE level (Fig. 2e). The optical response uses the full AO position matrices in addition to the Hamiltonian and overlap matrices. Figure 2e compares our method with the smaller Mo:2s1p1d, S:1s1p basis in the left panel and real-space reconstruction with the larger Mo:3s2p2d, S:3s2p1d basis in the right panel. Our method reproduces the dominant imaginary-part peak at 2.81 eV, coinciding with the QE reference, whereas the corresponding peak from real-space reconstruction occurs at 2.86 eV. Differences in spectral amplitudes remain for both representations.

For the displayed spectra, a constant real offset is fitted separately to each AO response over 0.01--5 eV: +16.84 Å for our method and +12.03 Å for real-space reconstruction. The imaginary parts and photon energies are unchanged. The complex normalized root-mean-square errors, evaluated at 500 equally weighted photon energies in this interval, are 7.99\% and 14.92\%, respectively, after this calibration. Before calibration, the corresponding errors are 13.76\% and 16.92\%. Thus, the displayed errors quantify agreement after fitting the real background. The comparison also tests transition matrix elements, extending the assessment beyond agreement in energy eigenvalues.

\begin{figure*}[tbp]
\centering
\begin{minipage}[t]{0.202\textwidth}
\raggedright\panelletter{a}\par\smallskip
\includegraphics[width=\linewidth,height=1.65909\linewidth]{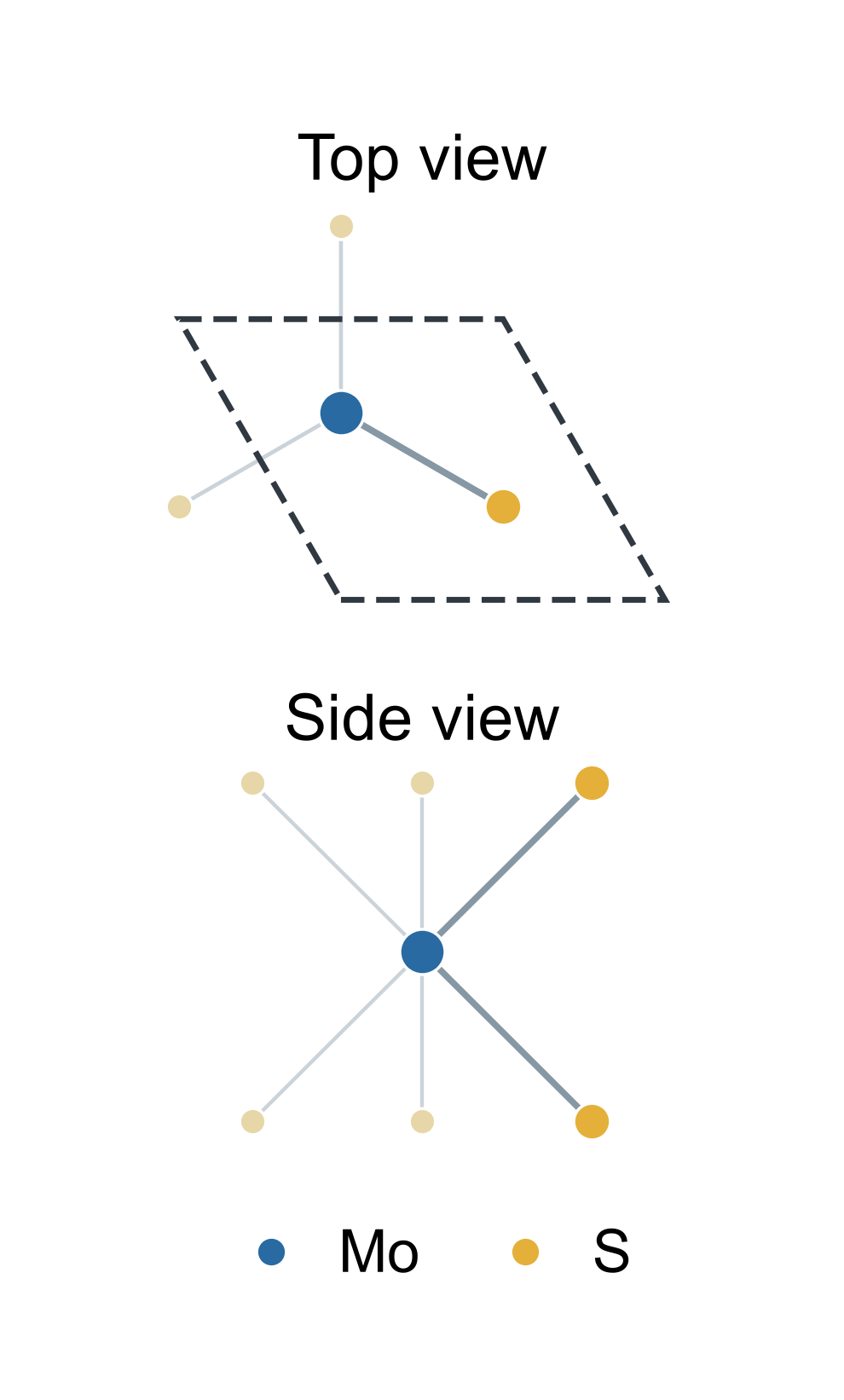}
\end{minipage}\hfill
\begin{minipage}[t]{0.335\textwidth}
\raggedright\panelletter{b}\par\smallskip
\includegraphics[width=\linewidth,height=1.0\linewidth]{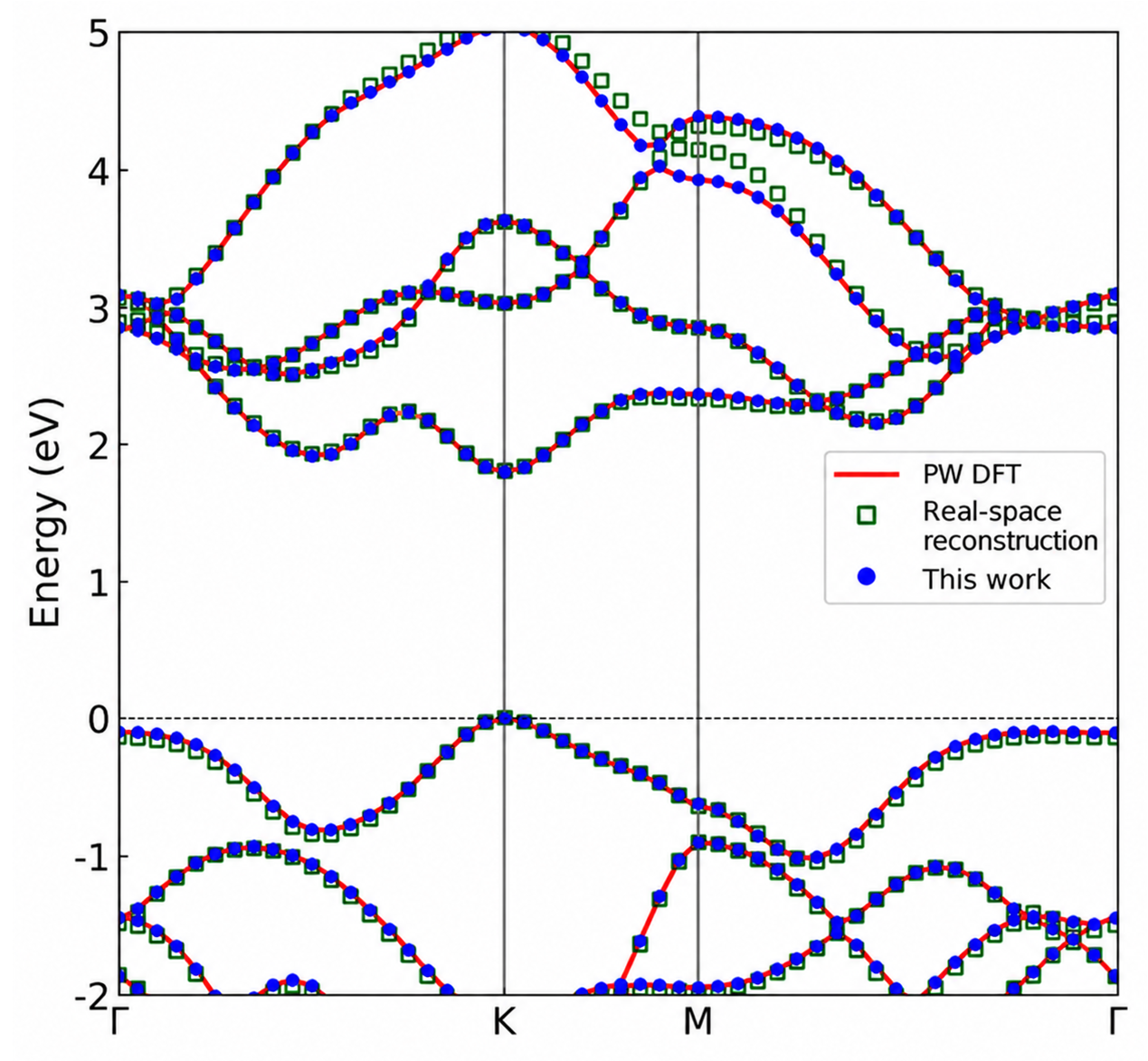}
\end{minipage}\hfill
\begin{minipage}[t]{0.335\textwidth}
\raggedright\panelletter{c}\par\smallskip
\includegraphics[width=\linewidth,height=1.0\linewidth]{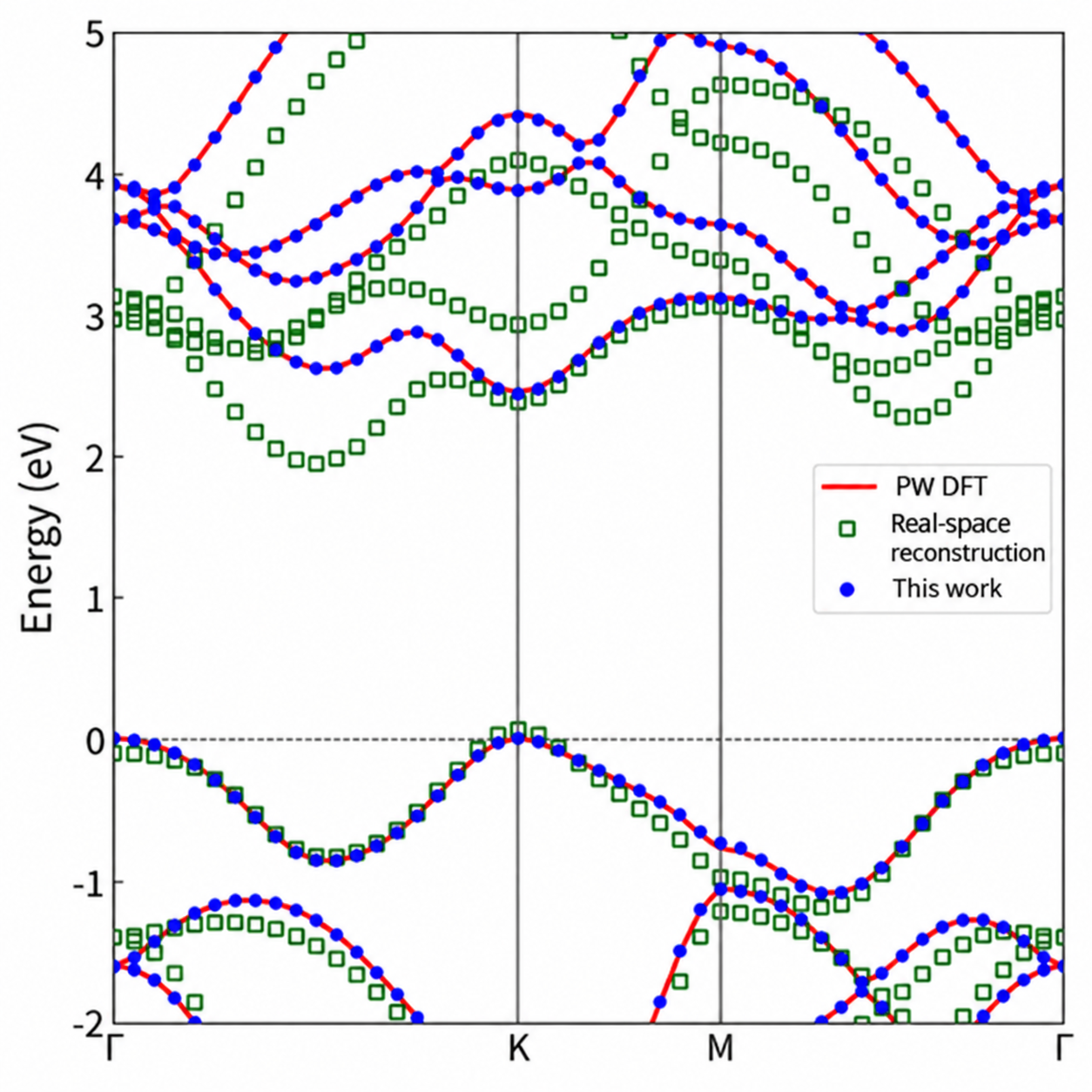}
\end{minipage}
\par\medskip
\begin{minipage}[t]{0.44\textwidth}
\raggedright\panelletter{d}\par\smallskip
\includegraphics[width=\linewidth]{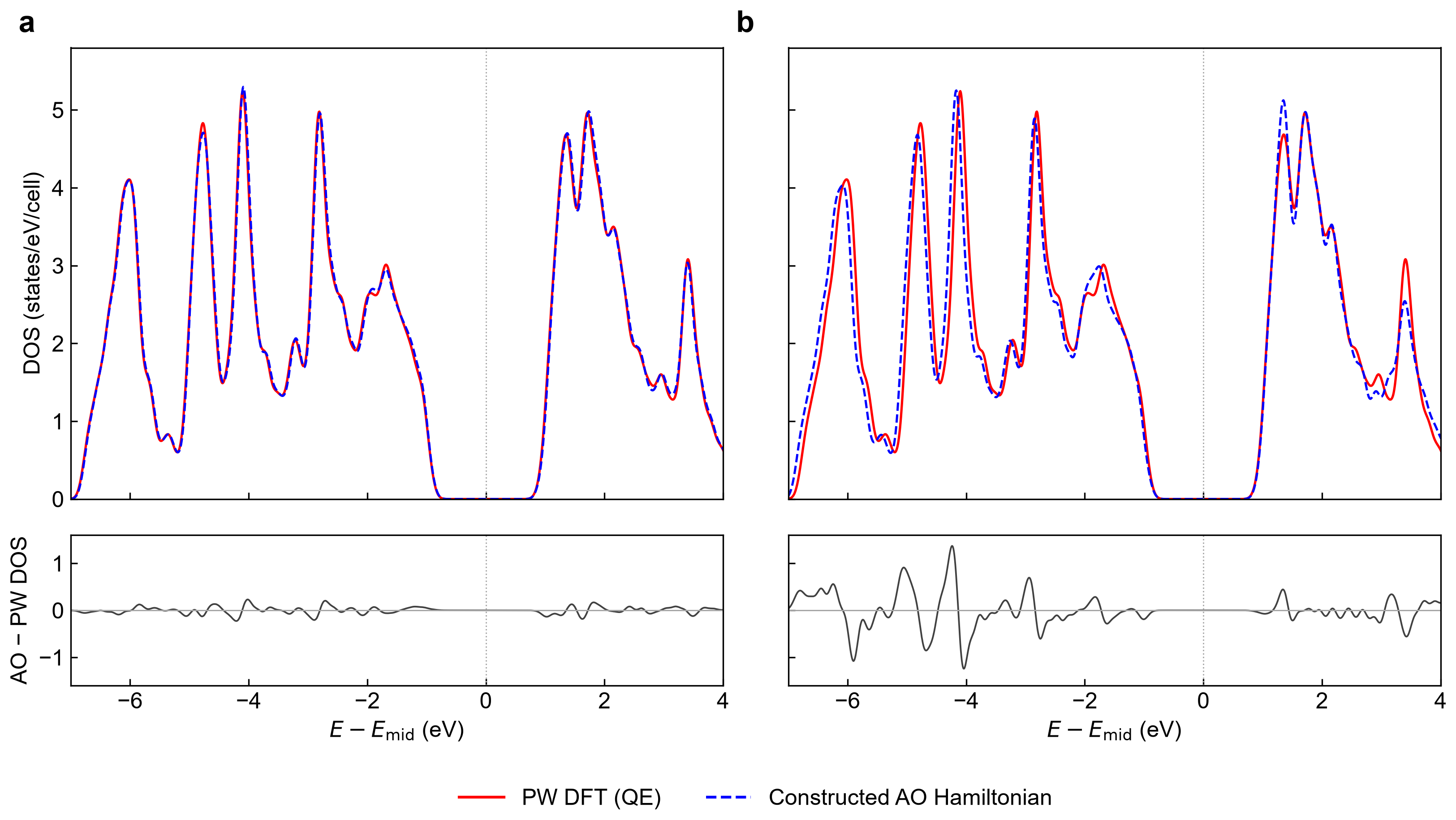}
\end{minipage}\hfill
\begin{minipage}[t]{0.44\textwidth}
\raggedright\panelletter{e}\par\smallskip
\includegraphics[width=\linewidth]{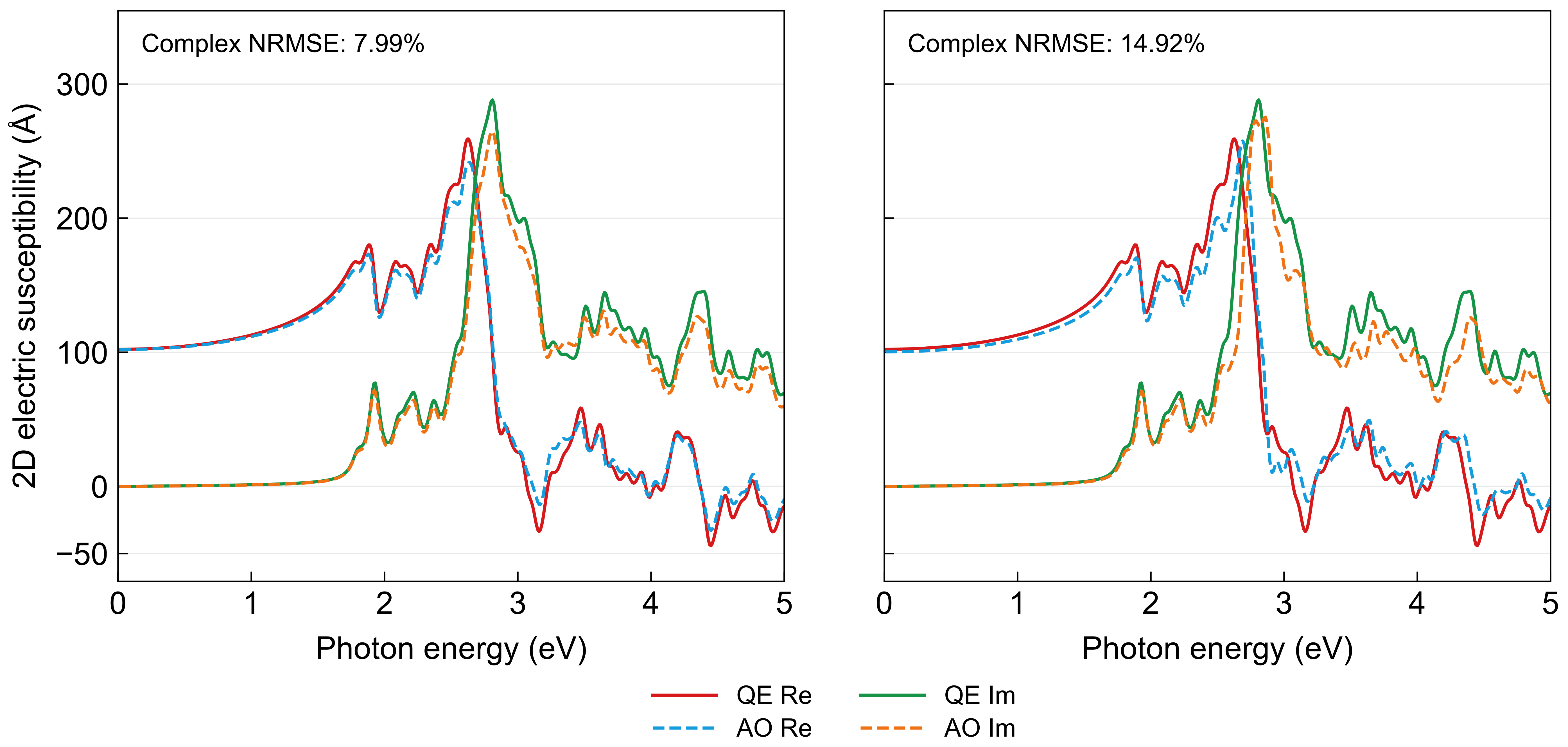}
\end{minipage}
\caption{\textbf{Structure, band structures, density of states and optical response of monolayer MoS\textsubscript{2}.} a, Top and side views of the primitive cell containing one Mo and two S atoms; blue and gold spheres represent Mo and S, and the dashed outline marks the cell. b, PBE band structures. c, HSE band structures along \ensuremath{\Gamma}--K--M--\ensuremath{\Gamma}; red lines show PW DFT, green squares real-space reconstruction with the larger Mo:3s2p2d, S:3s2p1d basis, and blue circles this work with the smaller Mo:2s1p1d, S:1s1p basis. d, Density of states (states/eV/cell) over \ensuremath{-}7 to 4 eV; the left and right panels show our construction and real-space reconstruction, respectively. Red solid and blue dashed curves show the PW DFT and AO DOS; lower plots show their differences. Each spectrum is aligned to its own midgap, with no additional shift or scaling. e, Complex electric susceptibility at the independent-particle PBE level; the left and right panels show our construction and real-space reconstruction, respectively, and both show the in-plane xx component. Solid red and green curves are the real and imaginary QE responses; dashed blue and orange curves are the AO responses. The real offsets (+16.84 and +12.03 Å, respectively) are fitted over 0.01--5 eV; the imaginary spectra and photon energies are unchanged. The plotted complex NRMSEs are evaluated after calibration using 500 equally weighted energy points.}
\label{fig:2}
\end{figure*}

We next compare the times required to construct AO Hamiltonians for monolayer MoS\textsubscript{2} supercells containing 6, 24, 54, 96 and 216 atoms (Fig. 3a). Our method uses the smaller Mo:2s1p1d, S:1s1p basis; real-space reconstruction is evaluated with both this basis and the larger Mo:3s2p2d, S:3s2p1d basis. Real-space reconstruction with the smaller basis shows approximately linear growth in construction time, consistent with the locality of the AO basis \cite{ref15,ref19,ref20,ref24}. Our method uses a restricted set of PW eigenstates, with k-point meshes chosen to maintain a consistent reciprocal-space spacing across supercells. For the tested 6--96-atom structures, our method is 2.11--3.37 times faster than real-space reconstruction with the same smaller basis. At 216 atoms, the respective times are 301.10 and 290.20 s, indicating that this advantage does not persist at the largest size for the same basis. Compared with real-space reconstruction using the larger basis, our method remains faster for all tested sizes, including 301.10 versus 475.00 s at 216 atoms. Thus, the computational advantage depends on both system size and basis choice; our construction does not exhibit linear scaling.

We also compare the diagonalization times of the AO Hamiltonians, normalized by the product of the number of bands and the number of k points (Fig. 3b). This comparison uses our method with the smaller Mo:2s1p1d, S:1s1p basis and real-space reconstruction with the larger Mo:3s2p2d, S:3s2p1d basis. For the tested 24--216-atom supercells, the smaller representation reduces the normalized diagonalization time by factors of 2.27--5.44. At 216 atoms, the values are 80 and 435 s\ensuremath{\cdot}(band\ensuremath{\cdot}k)\textsuperscript{\ensuremath{-}}\textsuperscript{1}, respectively, whereas both are 0.02 s\ensuremath{\cdot}(band\ensuremath{\cdot}k)\textsuperscript{\ensuremath{-}}\textsuperscript{1} at six atoms. These results quantify the benefit of the compact AO basis for the subsequent electronic-structure calculation.

\begin{figure*}[tbp]
\centering
\makebox[\linewidth][l]{\panelletter{a}\hspace{0.475\linewidth}\panelletter{b}}\par\smallskip
\includegraphics[width=0.9\linewidth]{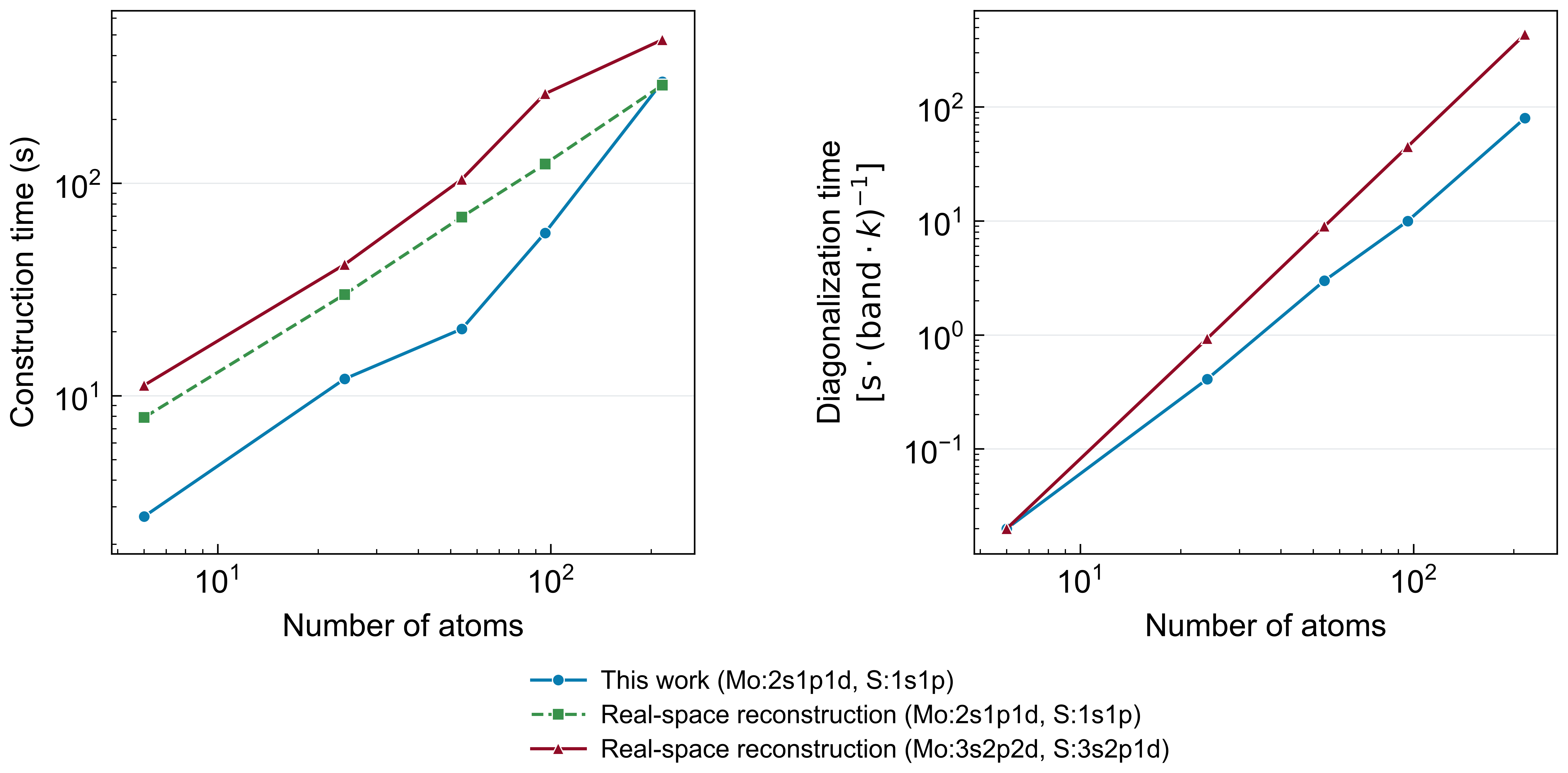}
\caption{\textbf{Hamiltonian construction and diagonalization times for MoS\textsubscript{2} supercells.} a, Construction time. Blue circles show our method using Mo:2s1p1d, S:1s1p. Green squares and burgundy triangles show real-space reconstruction using Mo:2s1p1d, S:1s1p and Mo:3s2p2d, S:3s2p1d, respectively. b, Diagonalization time normalized by the number of bands times the number of k points, in s\ensuremath{\cdot}(band\ensuremath{\cdot}k)\textsuperscript{\ensuremath{-}}\textsuperscript{1}. Blue circles and burgundy triangles compare our method with the smaller basis and real-space reconstruction with the larger basis. Both panels use logarithmic axes, and lines connect the measured values. The k-point meshes maintain a consistent reciprocal-space spacing across supercells.}
\label{fig:3}
\end{figure*}

\subsubsection*{Application to sodium phases under pressure}

Sodium phases under pressure provide a demanding test of compact AO representations \cite{ref35,ref36}. Compression changes orbital hybridization and can redistribute valence electrons into interstitial regions; in the insulating hP4 phase, p--d hybridization and repulsion from ionic cores produce pronounced interstitial electron localization \cite{ref37}. For self-consistent AO DFT, an atom-centered basis must represent these changes in both the radial and angular character of the electronic states. A basis suitable for one phase or pressure may therefore require radial optimization or additional basis functions to remain accurate in another \cite{ref17,ref31,ref32}. Maintaining this flexibility while keeping the basis compact is a central challenge, because enlarging the basis increases the Hamiltonian dimension and the cost of diagonalization.

We compare our construction with real-space reconstruction for bcc, fcc, cI16, oP8 and hP4 Na at the pressures examined. Our method uses a 2s2p1d basis, whereas real-space reconstruction uses a 4s3p2d basis, corresponding to 13 and 23 AO functions per atom, respectively. Thus, our construction reduces the basis dimension by approximately 43\%. The hP4 Na primitive cell is shown in Fig. 4a. Figure 4b compares the corresponding AO band structures with the PW DFT reference along \ensuremath{\Gamma}--M--K--\ensuremath{\Gamma}--A--L--H--A. As shown in Fig. 4c, the band-energy MAEs relative to PW DFT are 1, 3, 1, 1 and 1 meV for our method, compared with 12, 26, 14, 18 and 28 meV for real-space reconstruction, respectively. The accuracy advantage persists across all five phases. In particular, the error remains at 1 meV for hP4 Na, where real-space reconstruction gives the largest error of 28 meV despite its larger basis.

These results show that accurate reconstruction of the bands examined does not require a correspondingly large AO basis. By constructing the Hamiltonian from selected PW eigenstates and their eigenvalues, our method obtains a compact AO representation of the PW electronic reference. The small errors for cI16, oP8 and hP4 Na further demonstrate its accuracy across structurally and electronically distinct phases. The resulting reduction in Hamiltonian dimension offers an advantage for subsequent diagonalization.

\begin{figure*}[tbp]
\centering
\includegraphics[width=0.9\linewidth]{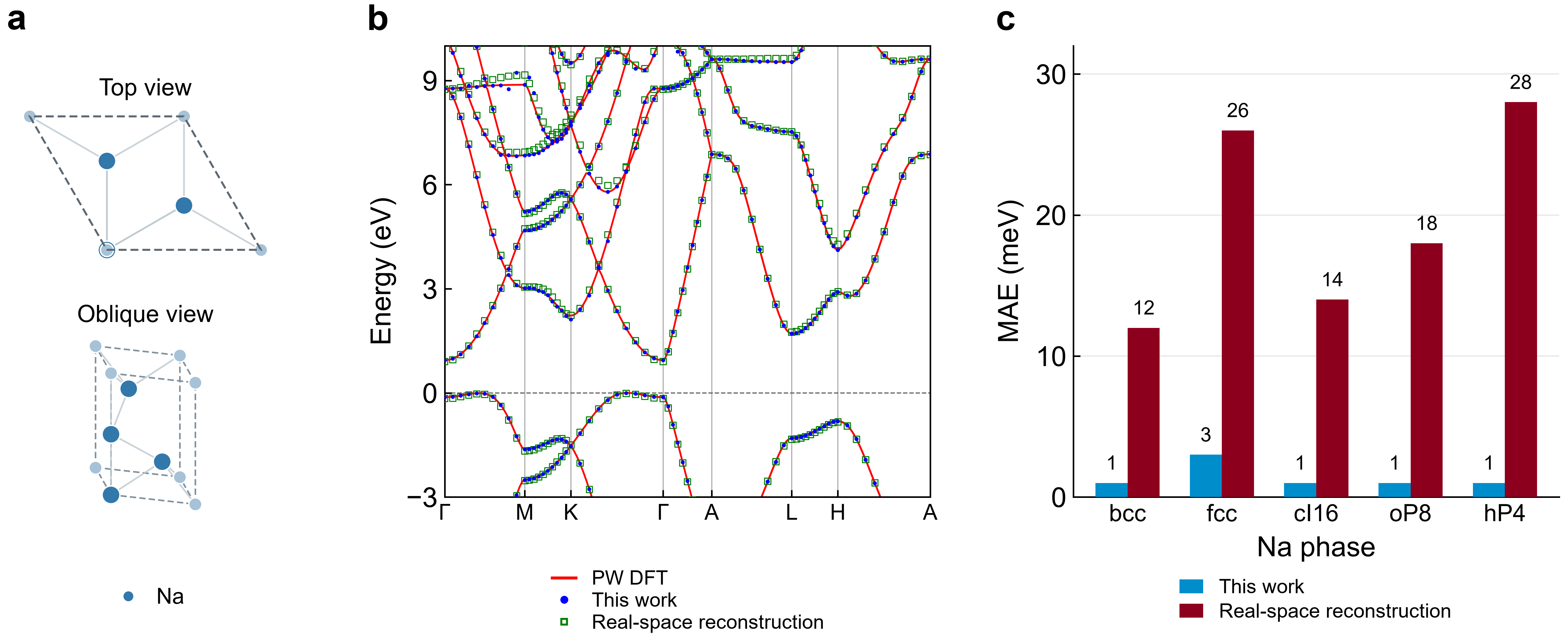}
\caption{\textbf{hP4 Na structure, band comparison and Hamiltonian reconstruction accuracy for sodium phases.} a, Top and oblique views of the hP4 Na primitive cell containing four Na atoms. Blue spheres represent Na atoms; lighter spheres show periodic equivalents at the cell boundaries, and dashed lines mark the cell. b, Constructed AO bands compared with PW DFT along \ensuremath{\Gamma}--M--K--\ensuremath{\Gamma}--A--L--H--A. Red solid lines denote PW DFT; blue circles denote our construction with the 2s2p1d basis, and green open squares denote real-space reconstruction with the 4s3p2d basis. Energies are referenced to the PW Fermi energy, with one global rigid alignment for each AO construction over its retained band set. c, Band-energy MAEs relative to PW DFT for bcc, fcc, cI16, oP8 and hP4 Na. Blue and red bars denote our method and real-space reconstruction, respectively, using 2s2p1d (13 functions per atom) and 4s3p2d (23 functions per atom) bases.}
\label{fig:4}
\end{figure*}

\subsubsection*{Application to Al surfaces}

We further assess Hamiltonian construction for a four-layer Al slab containing 16 atoms at the PBE level \cite{ref33}. Figure 5a shows the slab structure, and Fig. 5b,c compare the directly constructed AO band structures with the PW DFT reference along \ensuremath{\Gamma}--K--M--\ensuremath{\Gamma}. Our construction uses an Al:1s1p1d basis with 9 AO functions per atom, whereas real-space reconstruction uses an Al:3s3p2d1f basis with 29 functions per atom. The resulting AO Hamiltonian dimensions are 144 and 464, respectively, so our construction reduces the basis dimension by approximately 69\%.

After a single rigid alignment for each construction, fitted using the same PW reference points within \ensuremath{-}4 to 4 eV, the band-energy MAEs within \ensuremath{-}2 to 2 eV are 2.30 meV for our method and 11.55 meV for real-space reconstruction. Across the plotted \ensuremath{-}4 to 4 eV window, the corresponding MAEs are 3.01 and 40.82 meV. Our construction closely reproduces the PW dispersions near the Fermi level. The larger deviations of real-space reconstruction occur primarily above the reference energy zero: the MAEs over 0 to 4 eV are 4.31 and 69.27 meV, respectively. This test extends the compact-basis benchmark to a metallic slab. The comparison evaluates the chosen method and basis combinations, since the two constructions use different AO bases.

\begin{figure*}[tbp]
\centering
\includegraphics[width=0.9\linewidth]{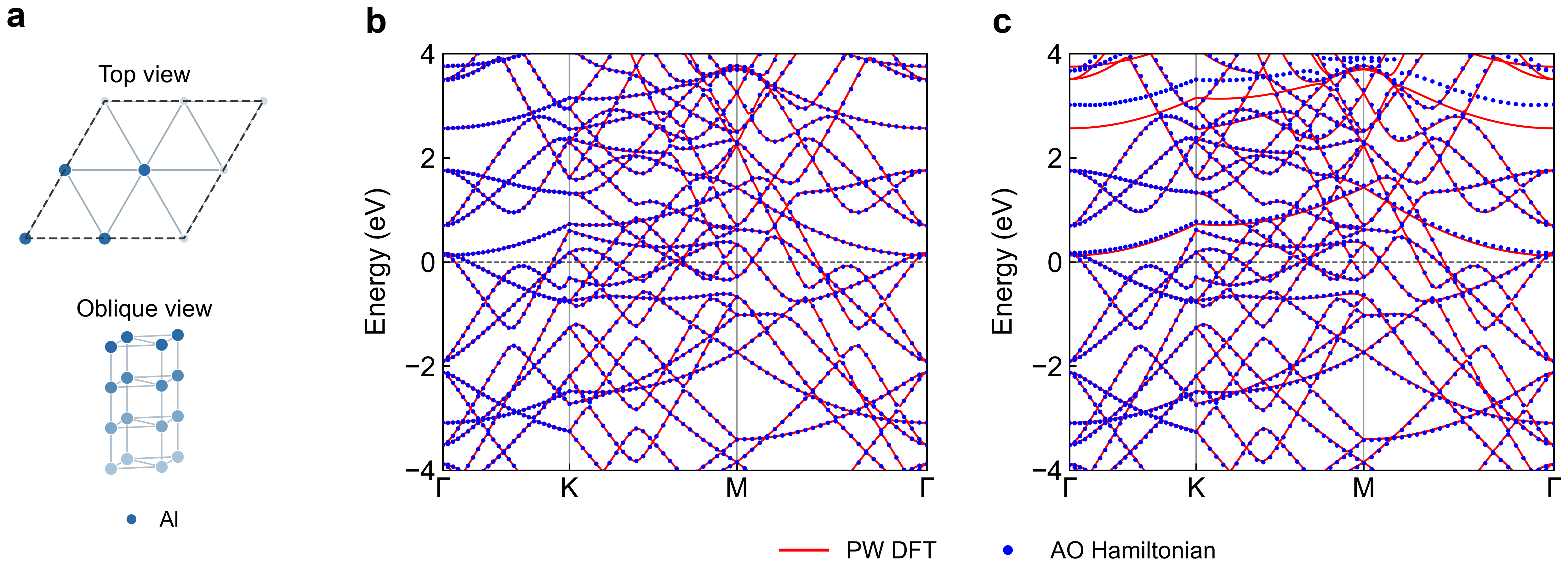}
\caption{\textbf{Al surface structure and constructed PBE band structures.} a, Top and oblique views of the 16-atom, four-layer Al slab used in the calculations. Blue spheres represent Al atoms, with lighter colours denoting lower layers and periodic neighbours. The dashed outline marks the in-plane cell. b, Our construction using the Al:1s1p1d basis, with 9 AO functions per atom. c, Real-space reconstruction using the Al:3s3p2d1f basis, with 29 AO functions per atom. Red solid lines show the PW DFT reference; blue circles in b and c show the directly constructed AO bands along \ensuremath{\Gamma}--K--M--\ensuremath{\Gamma}. Each construction is rigidly aligned using the same 4,181 PW-selected band--k points within \ensuremath{-}4 to 4 eV. This single alignment is used for the plot and all reported errors. The energy window is continuous because the reference bands cross the Fermi level.}
\label{fig:5}
\end{figure*}

\subsubsection*{Application to MgO with an oxygen vacancy}

We next examine a cubic MgO supercell containing one oxygen vacancy at the PBE level \cite{ref33}. The structure contains 32 Mg and 31 O atoms (Fig. 6a). Our construction uses Mg:2s2p and O:1s1p1d bases, whereas real-space reconstruction uses Mg:4s3p and O:3s3p2d bases. The AO Hamiltonian dimensions are 535 and 1098, respectively, corresponding to a reduction of approximately 51\% with our basis. Figures 6b,c compare the directly constructed AO bands with the PW DFT reference along \ensuremath{\Gamma}--X--M--\ensuremath{\Gamma}--R. The reference contains an isolated occupied band within the gap, separated from the lower occupied bands.

Using one rigid alignment for each construction, fitted to the same PW reference points within \ensuremath{-}3 to 3 eV, our method gives a band-energy MAE of 0.217 meV in this window, compared with 25.10 meV for real-space reconstruction. Across the plotted \ensuremath{-}5 to 5 eV window, the corresponding MAEs are 0.290 and 29.64 meV. For the isolated occupied band (band 253), the full-path MAEs are 0.205 and 110.62 meV, respectively. For the lowest unoccupied band (band 254), they are 0.149 and 63.11 meV. Our construction reproduces both the position and dispersion of the isolated in-gap band, while real-space reconstruction shifts this band upward relative to the PW reference. These results show that accurate reconstruction of isolated in-gap energy levels can be achieved with a compact AO basis.

\begin{figure*}[tbp]
\centering
\includegraphics[width=0.9\linewidth]{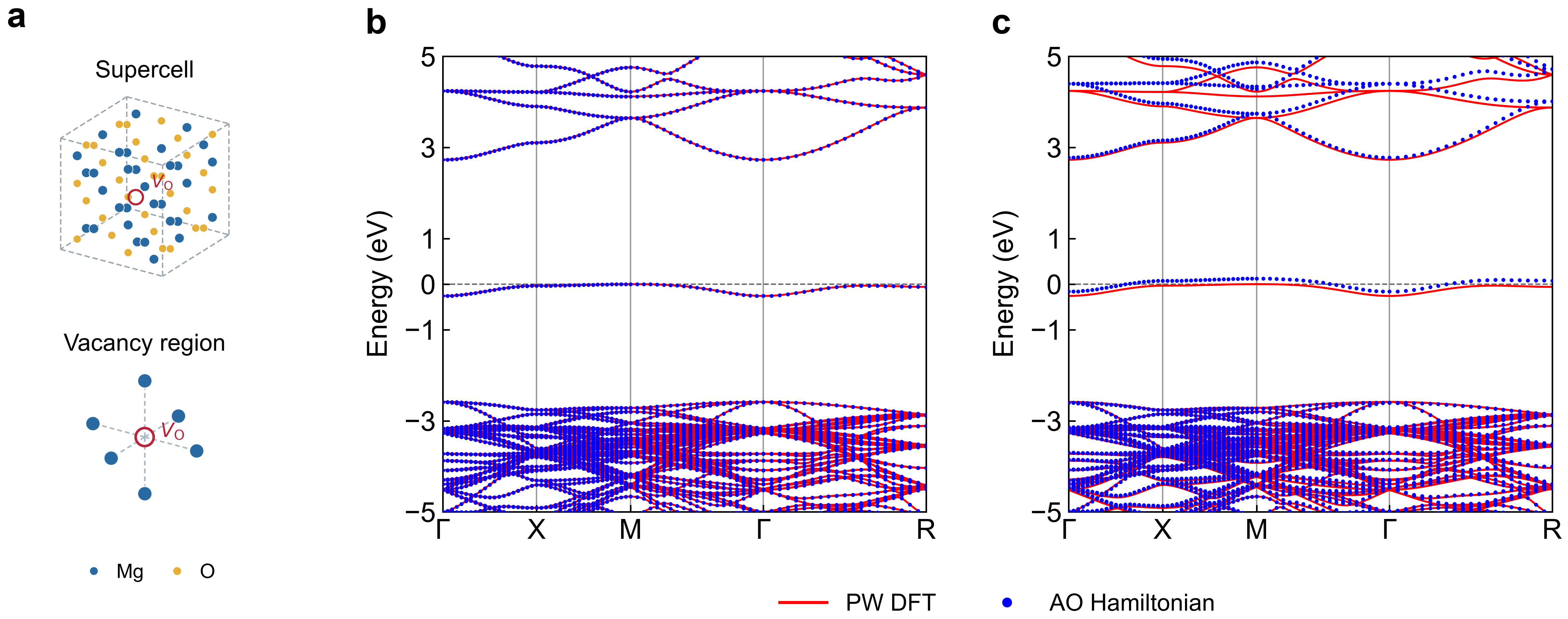}
\caption{\textbf{Oxygen-vacancy structure and constructed PBE bands of MgO.} a, The 63-atom MgO supercell and the local Mg environment surrounding the oxygen vacancy. Blue and gold spheres represent Mg and O, respectively. The red open circle marks the missing O site, and dashed lines indicate the supercell or local coordination. b, Our construction using Mg:2s2p and O:1s1p1d bases. c, Real-space reconstruction using Mg:4s3p and O:3s3p2d bases. Red solid lines show the PW DFT reference; blue circles in b and c show the constructed AO bands along \ensuremath{\Gamma}--X--M--\ensuremath{\Gamma}--R. Each construction is rigidly aligned using the same 1,012 PW-selected band--k points within \ensuremath{-}3 to 3 eV, and this alignment is used for the band panels and all reported errors. Band indices start from one.}
\label{fig:6}
\end{figure*}

\subsection*{Machine learning predictions}

The constructed real-space Hamiltonian matrices \(H(\mathbf{R})\) serve as training targets for an equivariant graph neural network, which predicts these matrices from atomic structures \cite{ref6,ref13}. The predicted Hamiltonians enable electronic-structure evaluation for new geometries without repeating self-consistent PW DFT calculations. Details of the equivariant neural network are provided in Methods.

\subsubsection*{Application to twisted bilayer MoS\textsubscript{2}}

Stacking and twist-induced structural rearrangements can strongly affect the electronic structure of bilayer MoS\textsubscript{2} \cite{ref34}. To assess the learnability of the constructed AO Hamiltonians, we generate a dataset of 1,150 non-twisted 3 \ensuremath{\times} 3 bilayer MoS\textsubscript{2} supercells with different stacking configurations, interlayer sliding and random atomic displacements. PW DFT calculations are performed using both PBE and HSE, and the resulting AO Hamiltonians are used to train separate models for the two functionals. No twisted structures are included in the training set. We then evaluate generalization to a bilayer MoS\textsubscript{2} supercell with a twist angle of 21.79\ensuremath{^{\circ}} and 42 atoms. The atomic structure is shown in Fig. 7a. Figures 7b and 7c compare the PBE bands predicted by models trained on AO Hamiltonians from our construction and real-space reconstruction with the PW DFT reference.

\begin{figure*}[tbp]
\centering
\includegraphics[width=0.9\linewidth]{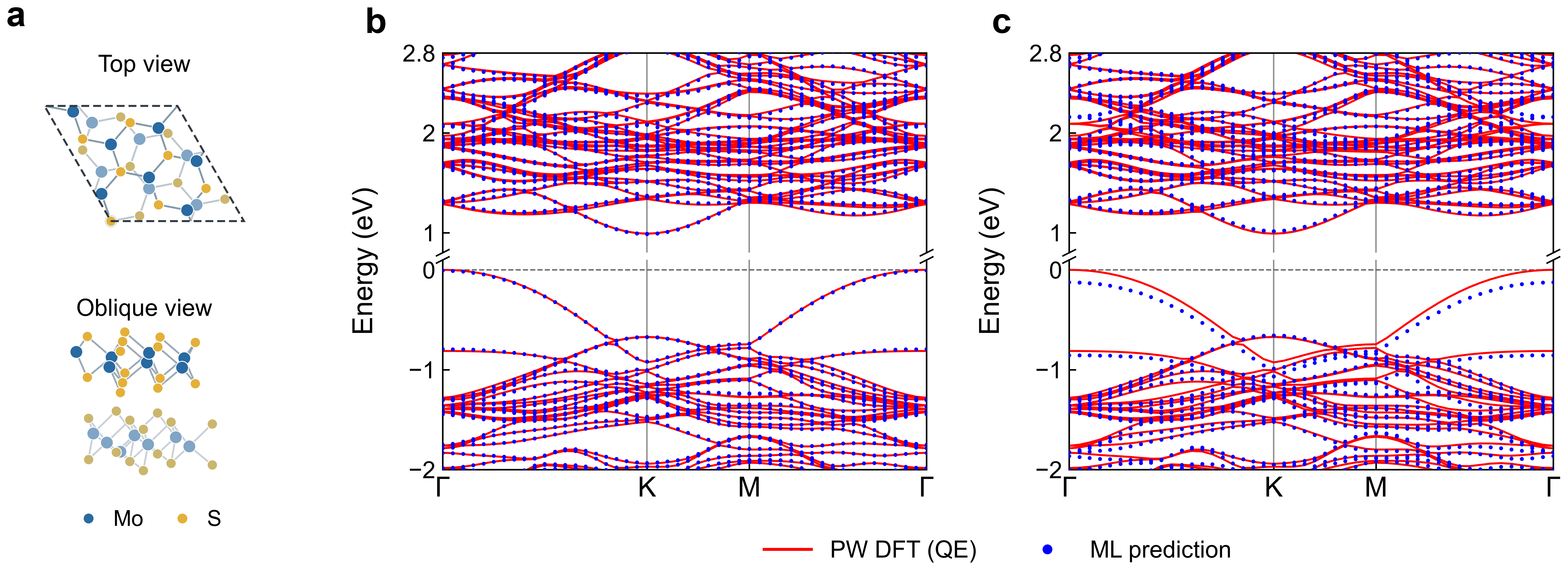}
\caption{\textbf{Twisted bilayer MoS\textsubscript{2} and predicted PBE band structures.} a, Top and oblique views of the 21.79\ensuremath{^{\circ}} crystal structure containing 42 atoms used for evaluation; blue and gold spheres represent Mo and S, respectively, with lighter colours denoting the lower layer. The dashed outline marks the periodic in-plane supercell. b, Model trained on AO Hamiltonians constructed using our method. c, Model trained on AO Hamiltonians obtained by real-space reconstruction. The two band panels show the predicted bands along \ensuremath{\Gamma}--K--M--\ensuremath{\Gamma}. Red lines denote PW DFT (QE); blue circles denote neural-network predictions. Our method uses the smaller Mo:2s1p1d, S:1s1p basis, whereas real-space reconstruction uses the larger Mo:3s2p2d, S:3s2p1d basis. The 0.1--0.8 eV interval is omitted for clarity. Each prediction is rigidly aligned using the \ensuremath{-}4 to 4 eV window, and the same alignment is used for the reported MAEs.}
\label{fig:7}
\end{figure*}

Within the plotted energy window of \ensuremath{-}2 to 2.8 eV, excluding the omitted interval of 0.1--0.8 eV, the predicted bands have MAEs of 7.71 meV for our method and 16.15 meV for real-space reconstruction. For the highest valence band, the MAEs along the full \ensuremath{\Gamma}--K--M--\ensuremath{\Gamma} path are 3.64 and 79.19 meV, respectively; for the lowest conduction band, they are 12.29 and 19.65 meV. Our model closely reproduces the highest valence band, whereas the model trained on real-space reconstruction exhibits a systematic deviation near \ensuremath{\Gamma}. These comparisons use the same PW DFT reference and a single rigid alignment for each prediction. The errors therefore assess the combined effects of Hamiltonian construction and neural-network prediction. The results demonstrate transfer from non-twisted training configurations to the tested twisted structure, with smaller band errors for the model trained on our constructed Hamiltonians.

\subsubsection*{Application to twisted bilayer graphene}

We further evaluate Hamiltonian learning for a twisted bilayer graphene supercell containing 28 atoms at the PBE level. Two sets of training AO Hamiltonians are generated from the same atomic structures using the same C:1s1p1d basis: one with our construction and the other with real-space reconstruction. Figure 8a shows the twisted structure, and Fig. 8b,c compare the band structures predicted by the resulting machine-learning models with the PW DFT reference along \ensuremath{\Gamma}--K--M--\ensuremath{\Gamma}. This comparison controls the structural and basis-set differences between the two training datasets and examines how the source of the training Hamiltonians affects the predicted electronic structure.

After a single rigid energy alignment for each prediction, the model trained on our Hamiltonians yields a band-energy MAE of 6.77 meV within \ensuremath{-}2 to 2 eV, compared with 19.28 meV for the model trained on real-space reconstruction. Across the plotted \ensuremath{-}4 to 4 eV window, the corresponding MAEs are 25.40 and 76.03 meV. For bands 56 and 57 near the Fermi level, the full-path MAEs are 21.15 and 6.63 meV for our training data, compared with 35.09 and 24.45 meV for real-space reconstruction. Both models reproduce the principal dispersions near K, while the model trained on our Hamiltonians gives closer overall agreement with PW DFT. A small 2.53 meV splitting between bands 56 and 57 remains at K in this prediction. These results support the accuracy benefit of our constructed Hamiltonians as training targets even when the two datasets use the same AO basis.

\begin{figure*}[tbp]
\centering
\includegraphics[width=0.9\linewidth]{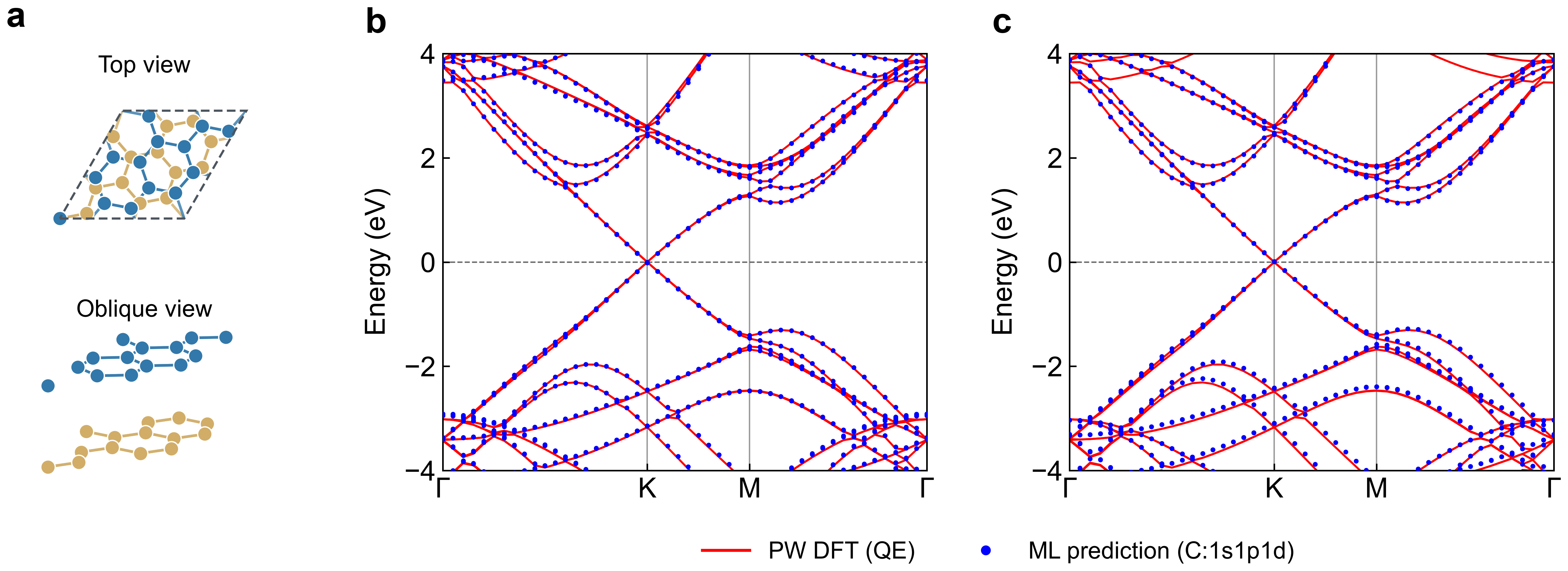}
\caption{\textbf{Structure and predicted PBE band structures of 28-atom twisted bilayer graphene.} a, Top and oblique views of the 21.79\ensuremath{^{\circ}} twisted bilayer graphene supercell. Blue and gold distinguish the two carbon layers, and the dashed outline marks the periodic cell in the top view. b, Machine-learning model trained on AO Hamiltonians constructed using our method. c, Machine-learning model trained on AO Hamiltonians obtained by real-space reconstruction. Both training datasets use the same atomic structures and the same C:1s1p1d AO basis. Red solid lines show the PW DFT (QE) reference; blue circles in b and c show the predicted bands along \ensuremath{\Gamma}--K--M--\ensuremath{\Gamma}. Each prediction is rigidly aligned using the same QE-defined \ensuremath{-}2 to 2 eV window, and this alignment is used for all reported errors.}
\label{fig:8}
\end{figure*}

\begin{figure*}[tbp]
\centering
\includegraphics[width=.44\textwidth]{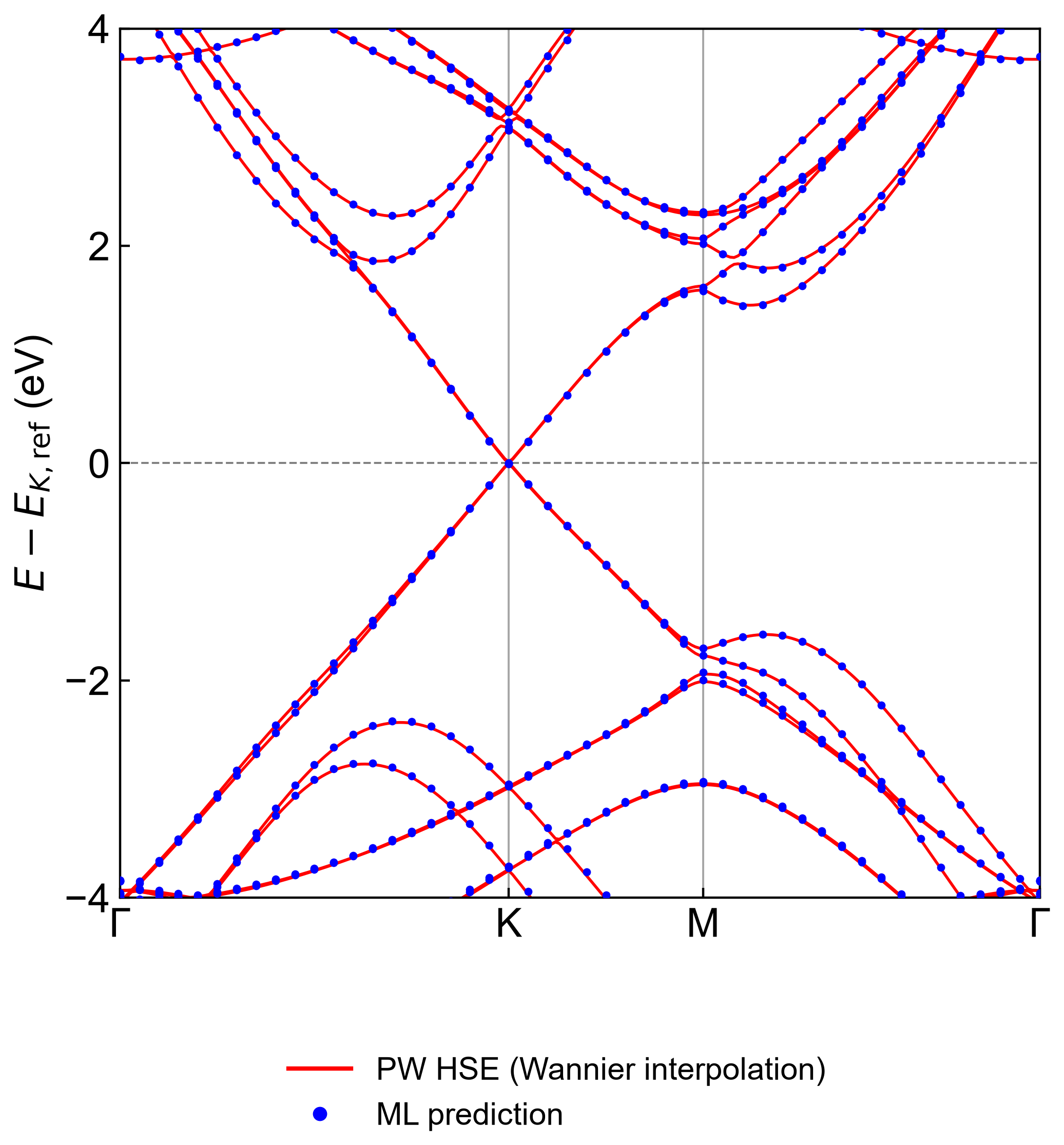}
\caption{\textbf{Predicted HSE band structure of 28-atom twisted bilayer graphene.} Red solid lines show the PW HSE reference obtained by Wannier interpolation; blue circles show the machine-learning prediction along \ensuremath{\Gamma}--K--M--\ensuremath{\Gamma}. The energy zero is the reference K-point midpoint of bands 56 and 57. A single rigid alignment obtained within \ensuremath{-}2 to 2 eV is used for the plot and all reported errors. Both \ensuremath{\Gamma} endpoints represent the same k point and use identical predicted eigenvalues.}
\label{fig:9}
\end{figure*}

We additionally evaluate the HSE \cite{ref26} band prediction for the 28-atom twisted bilayer graphene structure (Fig. 9). The PW HSE reference bands are obtained by Wannier interpolation. After a single rigid energy alignment, the predicted bands have MAEs of 7.42 meV within \ensuremath{-}2 to 2 eV and 11.04 meV within the plotted \ensuremath{-}4 to 4 eV window. The full-path MAEs for bands 56 and 57 are 7.79 and 9.89 meV, respectively. The model follows the principal HSE band dispersions, although the separation between bands 56 and 57 at K is 1.99 meV, compared with 11.41 meV in the Wannier-interpolated reference.

\section*{Discussion}

Our projectability-guided construction extends AO Hamiltonian learning to a broader range of PW DFT calculations, including hybrid functionals. By constructing the AO Hamiltonian from selected PW eigenstates and their eigenvalues, the method incorporates the effects of nonlocal contributions present in the reference calculation without explicitly evaluating their AO matrix elements. The complementary AO subspace is completed using selected higher-energy PW states, avoiding an empirical shift and its associated contribution to the real-space training targets. The resulting Hamiltonians can therefore connect established PW DFT workflows with equivariant Hamiltonian-learning frameworks.

The benchmarks demonstrate that accurate reproduction of the PW bands of interest can be achieved with compact AO bases. For the five sodium phases examined, our method yields band-energy MAEs of 1--3 meV using 13 AO functions per atom, compared with 12--28 meV for real-space reconstruction using 23 functions per atom. This reduction in Hamiltonian dimension lowers the cost of subsequent diagonalization. Although our construction does not exhibit linear scaling, it is faster than real-space reconstruction with the same smaller basis for the tested 6--96-atom MoS\textsubscript{2} supercells and remains faster than reconstruction with the larger basis at all tested sizes up to 216 atoms under the reciprocal-space sampling conditions used here. This efficiency is useful for generating training datasets from small and intermediate-sized structures. Training on non-twisted bilayer MoS\textsubscript{2} configurations and evaluating the resulting models on a twisted structure further illustrates how the constructed PBE and HSE Hamiltonians can be used as learning targets.

The accuracy of the construction remains tied to the ability of the AO basis to represent the selected PW states. Projectability provides a quantitative diagnostic of this representation and can guide basis enlargement or optimization of the radial functions. Avoiding an empirical shift removes one arbitrary element of the construction, but does not by itself establish the accuracy of neural-network predictions for unseen structures, which requires separate validation. Future work could combine projectability-guided basis optimization with broader structural datasets and extend the construction to other PW calculations containing additional nonlocal contributions. These developments would further expand the use of compact, machine-learned AO Hamiltonians across electronic-structure methods and materials.

\section*{Methods}

\subsection*{Hamiltonian construction details}

\textbf{Symmetry expansion and Fourier transform.} The Hamiltonian is constructed only at the IBZ k-points reported by Quantum ESPRESSO (QE) \cite{ref38,ref39}. Crystal space-group symmetries and time-reversal symmetry are used to expand to the full Monkhorst--Pack grid \cite{ref40} via covariant transformation:

\begin{equation}
H\left( g\mathbf{k} \right) = D_{g}\left( \mathbf{k} \right)H\left( \mathbf{k} \right){D_{g}\left( \mathbf{k} \right)}^{\dagger},
\label{eq:9}
\end{equation}

where \(D_{g}\left( \mathbf{k} \right)\) is the k-dependent AO representation matrix of symmetry operation \(g\), constructed from atomic site permutations and angular-momentum rotation matrices. A discrete inverse Bloch transform \cite{ref18} yields real-space matrices:

\begin{equation}
H\left( \mathbf{R} \right) = \frac{1}{N_{k}}\sum_{\mathbf{k}}{H\left( \mathbf{k} \right)}\exp\left( - i2\pi\mathbf{k} \cdot \mathbf{R} \right),
\label{eq:10}
\end{equation}

and analogously for \(S\left( \mathbf{R} \right)\). The Hermitian pairing \(H_{\beta\gamma}\left( \mathbf{R} \right) = {H_{\gamma\beta}\left( - \mathbf{R} \right)}^{\ast}\) is enforced, and boundary translations are handled according to the periodicity of the finite k-grid.

\subsection*{Preparation of datasets}

\textbf{Bilayer MoS\textsubscript{2} dataset.} We generate 1,150 non-twisted bilayer MoS\textsubscript{2} structures using 3 \ensuremath{\times} 3 supercells starting from 2H and 3R stackings. The structures include random in-plane interlayer sliding, random variations in the interlayer separation within \ensuremath{\pm}0.1 Å of the reference value, and random atomic displacements within \ensuremath{\pm}0.1 Å along each of the a, b and c directions. PW DFT calculations are performed using the PBE \cite{ref33} and HSE \cite{ref26} functionals with Quantum ESPRESSO \cite{ref38,ref39}. A 3 \ensuremath{\times} 3 \ensuremath{\times} 1 k-point mesh is used, with energy cutoffs of 70 Ry for the wavefunctions and 280 Ry for the charge density. AO Hamiltonian datasets are constructed from the PW results using our method and real-space reconstruction \cite{ref24}. The smaller AO basis is SZ. The larger basis is DZP augmented with one diffuse 4s function on sulfur with a cutoff radius of 8 a.u. Both numerical AO basis sets are generated using SIESTA \cite{ref15}.

\textbf{Bilayer graphene dataset.} We generate 600 bilayer graphene structures using 3 \ensuremath{\times} 3 supercells. Structural diversity is introduced through random in-plane interlayer sliding, variations in the interlayer separation within \ensuremath{\pm}0.1 Å of the reference value, and random atomic displacements within \ensuremath{\pm}0.1 Å along each of the a, b and c directions. The same 600 structures are used to prepare the PBE and HSE datasets. PW DFT calculations are performed with the PBE \cite{ref33} and HSE \cite{ref26} functionals using Quantum ESPRESSO \cite{ref38,ref39}, with a 5 \ensuremath{\times} 5 \ensuremath{\times} 1 k-point mesh for both functionals. The HSE calculations also use a 5 \ensuremath{\times} 5 \ensuremath{\times} 1 q-point mesh for exact exchange. AO Hamiltonians are constructed from the PW results using our method. A real-space reconstruction dataset \cite{ref24} is also prepared for the PBE comparison using the same structures and AO basis. The numerical AO basis functions are generated with SIESTA \cite{ref15}. The C:1s1p1d basis is used for both the PBE and HSE datasets.

\subsection*{Equivariant neural network details}

The constructed real-space matrices \(H\left( \mathbf{R} \right)\) serve as supervised training targets for an equivariant graph neural network that learns to predict \(H\left( \mathbf{R} \right)\) from crystal structure alone, enabling electronic-structure evaluation for new geometries without repeating the PW calculation.

\textbf{Periodic graph representation.} Each crystal structure is encoded as a periodic graph \cite{ref41}. Atoms correspond to nodes, characterised by element type, orbital configuration, and Cartesian coordinates. Edges connect atom pairs \((i,j)\) with lattice translation \(\mathbf{R} = \left( R_{x},R_{y},R_{z} \right)\); the associated displacement vector is \(\mathbf{d}_{ij}\left( \mathbf{R} \right) = \mathbf{r}_{j} + R_{1}\mathbf{a}_{1} + R_{2}\mathbf{a}_{2} + R_{3}\mathbf{a}_{3} - \mathbf{r}_{i}\). Edge distances are mapped to radial features via a learnable radial basis, and edge directions are expanded in real spherical harmonics up to order \(l_{\text{max}}\).

\textbf{SO(2)-equivariant message passing.} Message-passing networks aggregate information from neighbouring atoms \cite{ref42}. E(3)-equivariant networks represent angular information using spherical harmonics and Clebsch--Gordan (CG) tensor products \cite{ref43,ref44}. Efficient equivariant architectures have also been developed for Hamiltonian prediction \cite{ref45}. Following the SO(2) convolution strategy \cite{ref46}, the present network reduces angular complexity by constructing an edge-aligned local coordinate frame, rotating tensor features into that frame with Wigner-D matrices, applying learnable linear transformations within each magnetic quantum-number \(|m|\)-channel, and rotating back to the global frame. The message from atom \(j\) to atom \(i\) is:

\begin{equation}
m_{ij} = f_{\text{rad}}\left( r_{ij} \right) \odot {D_{ij}}^{- 1}\lbrack\bigoplus_{m}W_{m}\left( D_{ij}x_{ij} \right)_{m}\rbrack,
\label{eq:11}
\end{equation}

where \(x_{ij}\) is the input pair feature, \(D_{ij}\) is the global-to-local rotation, \(W_{m}\) is a learnable weight for channel \(|m|\), \(f_{\text{rad}}\left( r_{ij} \right)\) is a distance-dependent radial weight, and \(\odot\) denotes element-wise multiplication. Node features are updated by aggregating incoming messages and applying a nonlinear update, preserving equivariance under SO(3) rotations at reduced angular complexity.

\textbf{Euclidean Fast Attention (EFA).} Standard message passing is restricted to a finite cutoff radius and cannot efficiently capture long-range structural correlations. The EFA module \cite{ref47} addresses this by projecting atomic positions onto a set of \(N_{g}\) sampled directions on the unit sphere, applying Euclidean rotary positional encoding (RoPE) \cite{ref47,ref48} to query and key features, and aggregating information through an equivariant tensor-product attention:

\begin{equation}
h_{i}^{EFA} = \sum_{g = 1}^{N_{g}}w_{g}\lbrack Y\left( \widehat{u_{g}} \right) \otimes \left( \frac{1}{Z_{i}}\sum_{j \in \mathcal{N}_{i}}{\langle{q_{i}}^{(g)},{k_{j}}^{(g)}\rangle}v_{j} \right)\rbrack,
\label{eq:12}
\end{equation}

where \({q_{i}}^{(g)}\), \({k_{j}}^{(g)}\), and \(v_{j}\) are query, key, and value features incorporating direction-dependent positional encoding; \(Y\left( \widehat{u_{g}} \right)\) are spherical harmonics evaluated at grid direction \(\widehat{u_{g}}\); \(\otimes\) is an equivariant tensor product; \(w_{g}\) are quadrature weights; and \(Z_{i}\) is a normalisation factor. Because the sums over atoms and over directions factorise, the attention cost scales linearly with atom count at fixed direction-grid size \cite{ref47}.

The implementation supports both Fibonacci-sphere and lattice-derived direction grids. For periodic systems, local periodic images within a specified cutoff radius are included in the attention neighbourhood.

\textbf{Network architecture and output.} The full network consists of \(N_{sr}\) short-range SO(2) message-passing blocks followed by \(N_{lr}\) long-range EFA blocks. After the final block, an output module maps the updated node- and edge-level features to Hamiltonian matrix blocks \({\widehat{H}}_{ij}\left( \mathbf{R} \right)\) of dimension \(n_{i} \times n_{j}\), matching the orbital configuration of each atom pair.

\textbf{Evaluation of predicted MoS\textsubscript{2} bands.} Each prediction is aligned to the PW DFT reference using a single rigid shift that minimizes the absolute energy differences for the first 252 bands within a \ensuremath{-}4 to 4 eV window, selected using the predicted energies before alignment. The same shift is then used for every reported error. The MAE in the plotted window is evaluated at points whose PW DFT reference energies lie between \ensuremath{-}2 and 2.8 eV, excluding 0.1--0.8 eV; this gives the same 2,860 band--k points for both models. Highest-valence-band and lowest-conduction-band MAEs are averaged over all 48 points along \ensuremath{\Gamma}--K--M--\ensuremath{\Gamma}, using bands 182 and 183, respectively, with band indices starting from one.

\textbf{Evaluation of predicted graphene bands.} For the 28-atom twisted bilayer graphene structure, each prediction is aligned by a single rigid shift minimizing the absolute energy differences within the QE-defined \ensuremath{-}2 to 2 eV window among the first 62 bands. The same 186 band--k points are used for both alignments. Window MAEs are evaluated using QE energies to select common reference points: 186 points for \ensuremath{-}2 to 2 eV and 796 points for \ensuremath{-}4 to 4 eV. The wider window includes bands 44--65. Individual MAEs for bands 56 and 57 use all 48 k points along \ensuremath{\Gamma}--K--M--\ensuremath{\Gamma}. Band indices start from one.

\textbf{Evaluation of HSE graphene bands.} The HSE prediction contains 48 k points and 112 bands; the PW HSE reference contains 238 k points and is obtained by Wannier interpolation. Energies are ordered at each k point, and the reference is interpolated piecewise linearly along the same \ensuremath{\Gamma}--K--M--\ensuremath{\Gamma} path onto the prediction points. The reference K-point midpoint of bands 56 and 57, \ensuremath{-}0.564648 eV, defines the energy zero. A single rigid shift is fitted on 110 reference-selected band--k points within \ensuremath{-}2 to 2 eV among the first 62 bands. The same alignment is used for the \ensuremath{-}4 to 4 eV MAE, evaluated on 548 reference-selected points. The repeated terminal \ensuremath{\Gamma} block is replaced by the initial \ensuremath{\Gamma} block before analysis. Individual band MAEs include all 48 path points.

\textbf{Al surface calculations and band comparison.} PW DFT calculations for the 16-atom, four-layer Al slab use the PBE functional \cite{ref33} and Quantum ESPRESSO \cite{ref38,ref39}. The wavefunction and charge-density cutoffs are 80 and 640 Ry, respectively. Marzari--Vanderbilt smearing is used with a width of 0.02 Ry, and the out-of-plane supercell length is 28 Å. The \ensuremath{\Gamma}--K--M--\ensuremath{\Gamma} path contains 151 k points. The PW reference contains 150 bands, while our construction and real-space reconstruction yield 144 and 464 AO bands, respectively. Band energies are compared by ordered band index on the verified common k path. For each construction, a single L1-optimal rigid energy offset is determined from the same 4,181 PW-selected points within \ensuremath{-}4 to 4 eV among the first 144 reference bands. This offset is applied unchanged when evaluating narrower energy windows. The 0.1--0.8 eV interval is retained in this metallic system.

\textbf{MgO oxygen-vacancy calculations and band comparison.} PW DFT calculations use the PBE functional \cite{ref33} and Quantum ESPRESSO \cite{ref38,ref39} for a cubic 63-atom supercell with a lattice length of 8.424 Å, containing 32 Mg atoms and 31 O atoms. The wavefunction and charge-density cutoffs are 80 and 640 Ry, respectively. The structure used for the comparison is taken directly from the supplied calculation input. The missing oxygen site is identified by periodic matching of the atomic coordinates to the ideal O sublattice; its fractional coordinates are (0.25, 0.25, 0.25). The \ensuremath{\Gamma}--X--M--\ensuremath{\Gamma}--R path contains 161 k points. The supplied PW reference contains 900 bands, while the two AO constructions yield 535 and 1098 bands. A single L1-optimal energy offset is fitted independently for each construction on the same 1,012 PW-selected points within \ensuremath{-}3 to 3 eV among the first 535 common band indices. The same offset is used for the \ensuremath{-}5 to 5 eV comparison and all individual band errors. The reported MAEs for bands 253 and 254 include all 161 path points. Their assignment as the isolated highest occupied band and lowest unoccupied band is based on the PW reference spectrum.

\textbf{Density-of-states calculations.} The AO DOS is calculated on a 24 \ensuremath{\times} 24 \ensuremath{\times} 1 k-point mesh using Gaussian broadening with a standard deviation of 0.08 eV and a spin degeneracy of two. The occupied subspace contains 13 bands. Each AO spectrum is shifted to the midpoint of its eigenvalue gap; the QE spectrum uses a midgap estimated from its occupied-state plateau. The QE DOS is interpolated onto the common AO energy grid without additional broadening. The normalized root-mean-square error is evaluated over \ensuremath{-}7 to 4 eV and normalized by the root-mean-square PW DOS in the same interval. No fitted energy shift, background correction or amplitude rescaling is applied.

\textbf{Optical-response calculations.} The complex in-plane susceptibility is evaluated within the independent-particle approximation using full AO position matrices and the nonorthogonal Hamiltonian and overlap matrices. Velocity matrix elements include Hamiltonian derivatives, overlap derivatives and position-matrix contributions \cite{ref49}. The response kernel matches that used by QE epsilon.x. Both AO calculations use a 24 \ensuremath{\times} 24 \ensuremath{\times} 1 k mesh, a broadening parameter of 0.10 eV and a supercell height of 20 Å. The conductivity is converted to a two-dimensional electric susceptibility using the supercell height. For Fig. 2e, a constant real offset is fitted independently for each method on 500 points from 0.01 to 5 eV with 0.01 eV spacing. The imaginary responses and energy grids are preserved. The complex error combines the squared real- and imaginary-part differences and is normalized by the squared norm of the QE reference; no static AO response is inferred at zero frequency.

\end{document}